\documentclass{article}
\usepackage{spconf,amsmath,graphicx,hyperref}
\usepackage{amssymb}
\usepackage{algorithmic}
\usepackage{textcomp}
\usepackage{xcolor}
\usepackage{booktabs, tabularx}
\usepackage{multirow}
\usepackage[font=small]{caption}

\title{DBFM: Dual-Band Flow Matching for ECG Super-Resolution}
\name{Fan-Yi Hsu$^{1,2}$, Kuan-Chen Wang$^{2,3}$, I Chiu$^{2}$, Kai-Chun Liu$^{1}$, Ping-Cheng Yeh$^{3}$, and Yu Tsao$^{2}$}
\address{
$^{1}$National Yang Ming Chiao Tung University, 
$^{2}$Academia Sinica,
$^{3}$GICE, National Taiwan University \\
\small {Email: fanyi194.cs12@nycu.edu.tw, yu.tsao@citi.sinica.edu.tw}
}
\begin{document}
\ninept
\maketitle

\begin{abstract}
Long-term ECG monitoring is increasingly supported by wearable and portable devices, where limited power and transmission capacity motivate acquisition at lower sampling rates. ECG super-resolution (SR) aims to reconstruct high-sampling-rate signals from low-sampling-rate recordings for clinical analysis. A recent diffusion-based approach, DiSR-ECG, has demonstrated powerful reconstruction performance, particularly in cross-dataset evaluation. However, its iterative sampling during inference incurs high computational cost when processing large volumes of long-term ECG data. To reduce this cost while maintaining reconstruction accuracy, we propose Dual-Band Flow Matching (DBFM), a generative framework for ECG SR. DBFM decomposes ECG signals into low- and high-frequency components and separately learns their source-to-target transport using conditional flow matching. This dual-band formulation enables frequency-specific reconstruction, while one-step inference substantially reduces computational complexity. Experimental results show that DBFM achieves performance comparable to or better than DiSR-ECG while reducing inference FLOPs by approximately 98.7\%. These results demonstrate that DBFM enables computationally efficient ECG SR without compromising reconstruction quality.

\end{abstract}
\section{Introduction}
Cardiovascular diseases remain the leading cause of death worldwide, underscoring the importance of continuous and long-term monitoring of cardiac activity~\cite{roth2020global}. Electrocardiogram (ECG) signals provide a noninvasive measure of cardiac electrical activity and are widely used for cardiovascular diagnosis and monitoring~\cite{KAPLANBERKAYA2018216}. Wearable and portable devices have further extended ECG monitoring beyond conventional clinical settings~\cite{bayoumy2021smart, 8300311}. To support long-term ECG monitoring on devices with limited power and transmission budgets, one strategy is to record ECG at a low sampling rate and subsequently reconstruct high-sampling-rate signals on a more powerful platform, such as a cloud server or personal computer~\cite{8351558,bregar2022power,pizzuti1985digital}. In this framework, ECG super-resolution (SR) aims to reconstruct high-resolution (HR) ECG signals from their low-resolution (LR) counterparts, thereby recovering fine-grained ECG morphology for downstream analysis.

Contemporary ECG SR methods have mainly adopted neural networks for precise ECG reconstruction~\cite{srecg,dcaesr,msecg,disrecg}. Among them, DiSR-ECG~\cite{disrecg} employs residual-shifting diffusion with a Mamba-based architecture and demonstrates strong ECG reconstruction performance, particularly in cross-dataset evaluation against single-pass regression-based ECG SR approaches. However, diffusion-based inference requires repeated model evaluations during iterative sampling, resulting in substantial computational overhead~\cite{yue2023resshift}. This burden becomes increasingly pronounced for long-term ECG monitoring, where large volumes of data must be reconstructed. Hence, achieving accurate and robust ECG reconstruction at a lower inference cost remains an important challenge.

Flow matching~\cite{lipman2022flow} provides an efficient framework for generative modeling by learning a continuous vector field that transports samples from a source distribution toward a target distribution. Conditional flow matching learns this time-dependent vector field from samples along a prescribed probability path. Compared with conventional diffusion sampling, flow matching can enable efficient generation with fewer numerical integration steps. Recent studies have demonstrated its effectiveness for audio super-resolution~\cite{yun2025flowhigh,choi2026universr} and ECG-related tasks, including PPG-to-ECG translation~\cite{10888856} and ECG generation~\cite{bondar2025flowecg,11464731}, suggesting its potential for efficient generative ECG reconstruction.

Another potential limitation of existing ECG SR frameworks is that they reconstruct the entire ECG signal without separately processing its frequency components \cite{srecg,dcaesr,msecg}. This design may overlook the frequency-dependent effects of downsampling. Lower-frequency waveform structures are generally better preserved in the LR signal, while higher-frequency components undergo greater distortion and information loss. Moreover, because ECG energy is concentrated predominantly at lower frequencies~\cite{tereshchenko2015frequency}, we hypothesize that full-band optimization may be dominated by lower-frequency components that are already relatively well preserved, potentially limiting the reconstruction of fine-scale morphological details.

To address these challenges, we propose Dual-Band Flow Matching (DBFM), a flow-matching framework for robust ECG SR. DBFM decomposes ECG signals into low- and high-frequency components and learns the source-to-target transport of each band separately using conditional flow matching. By combining one-step flow-matching inference with dual-band reconstruction, DBFM enables frequency-specific reconstruction while maintaining substantially lower computational cost than the diffusion-based DiSR-ECG. Experiments on the PTB-XL and Chapman-Shaoxing ECG datasets show that DBFM significantly outperforms DiSR-ECG on three of four reconstruction metrics in both in-domain and cross-dataset evaluations, with no significant difference on the remaining metric. Moreover, DBFM requires only two function evaluations and reduces FLOPs by approximately 98.7\% compared with DiSR-ECG.

\section{Related work}
\begin{figure*}[t]
    \centering
    \includegraphics[width=.9\textwidth]{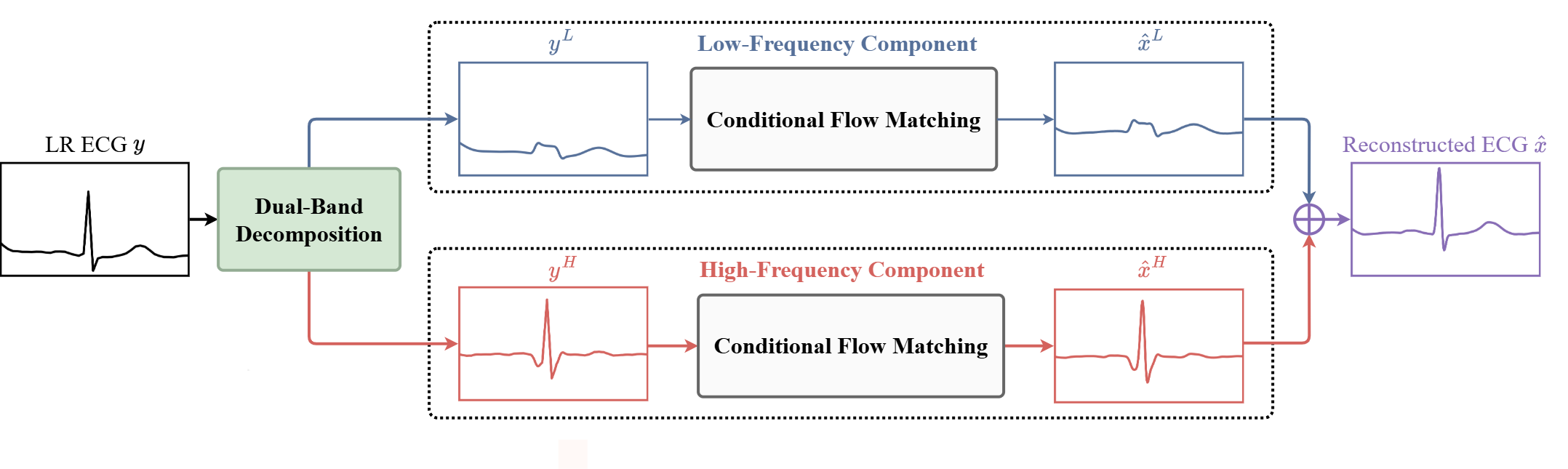}
    \caption{Overview of the proposed framework.}
    \label{fig:architecture}
\end{figure*}

\subsection{ECG super-resolution approaches}

ECG SR aims to reconstruct high-resolution ECG signals from their low-resolution counterparts. Conventional methods, such as linear interpolation (LI), estimate missing samples from neighboring observations but have limited capability to recover fine-grained waveform details. More recently, neural-network-based approaches have been developed, primarily based on direct regression. SRECG~\cite{srecg} employs convolutional residual blocks for ECG reconstruction, DCAE-SR~\cite{dcaesr} adopts a denoising convolutional autoencoder to jointly perform signal reconstruction and noise suppression, and MSECG~\cite{msecg} incorporates Mamba for ECG SR. In addition to direct regression, DiSR-ECG~\cite{disrecg} introduces a generative approach based on residual-shifting diffusion with a Mamba backbone and demonstrates strong reconstruction performance, particularly in cross-dataset evaluation. However, the inference stage of diffusion models relies on iterative sampling over multiple steps, resulting in substantially higher computational cost and inference latency than single-pass regression approaches.

\subsection{Flow matching}

Flow Matching (FM)~\cite{lipman2022flow} provides a simulation-free framework for training continuous normalizing flows~\cite{chen2018neural}. Let $\mathbf{x}\in\mathbb{R}^d$ denote a sample and $t\in[0,1]$ denote continuous time. FM considers a time-dependent probability path $p_t$ connecting a source distribution $p_0$ to a target distribution $p_1$. The path is generated by a flow $\phi_t:\mathbb{R}^d\rightarrow\mathbb{R}^d$ governed by a time-dependent vector field $v_\theta:\mathbb{R}^d\times[0,1]\rightarrow\mathbb{R}^d$:
\begin{equation}
    \frac{d}{dt}\phi_t(\mathbf{x})
    =
    v_\theta\!\left(\phi_t(\mathbf{x}),t\right),
    \qquad
    \phi_0(\mathbf{x})=\mathbf{x}.
\end{equation}

Conditional Flow Matching (CFM) constructs the marginal probability path from conditional distributions $p_t(\mathbf{x}\mid\mathbf{z})$, where $\mathbf{z}$ indexes the conditional path. A Gaussian conditional path is given by
\begin{equation}
    p_t(\mathbf{x}\mid\mathbf{z})
    =
    \mathcal{N}\!\left(
        \mathbf{x};
        \boldsymbol{\mu}_t(\mathbf{z}),
        \sigma_t(\mathbf{z})^2\mathbf{I}
    \right),
\end{equation}
where $\boldsymbol{\mu}_t(\mathbf{z})$ and $\sigma_t(\mathbf{z})$ denote the time-dependent mean and standard deviation, respectively. The corresponding conditional vector field is
\begin{equation}
    u_t(\mathbf{x}\mid\mathbf{z})
    =
    \frac{\sigma'_t(\mathbf{z})}{\sigma_t(\mathbf{z})}
    \left(
        \mathbf{x}-\boldsymbol{\mu}_t(\mathbf{z})
    \right)
    +
    \boldsymbol{\mu}'_t(\mathbf{z}),
\end{equation}
where the prime denotes differentiation with respect to $t$.
The vector field is learned by minimizing the CFM objective
\begin{equation}
    \mathcal{L}_{\mathrm{CFM}}
    =
    \mathbb{E}_{t,q(\mathbf{z}),p_t(\mathbf{x}\mid\mathbf{z})}
    \left[
        \left\|
        v_\theta(\mathbf{x},t)
        -
        u_t(\mathbf{x}\mid\mathbf{z})
        \right\|^2
    \right],
    \label{eq:cfm}
\end{equation}
with $t\sim\mathcal{U}(0,1)$.

\section{Proposed Method}

We propose DBFM, an efficient flow-matching framework for ECG super-resolution, as illustrated in Fig.~\ref{fig:architecture}. DBFM decomposes the ECG into low- and high-frequency components and applies conditional flow matching to each component for single-step reconstruction.

\subsection{Frequency decomposition}

ECG signals contain waveform structures at different temporal scales, ranging from slowly varying components to rapid morphological variations~\cite{4121752}. Motivated by the series decomposition scheme in Autoformer~\cite{wu2021autoformer}, we apply a moving-average decomposition to both the LR observation and HR target. Given an ECG signal $\mathbf{x}\in\mathbb{R}^{C\times T}$, the low-frequency component is obtained by smoothing the signal, while the high-frequency component is defined as the residual:
\begin{equation}
    \mathbf{x}^{L}
    =
    \operatorname{MovingAvg}_{K}(\mathbf{x}), \qquad
    \mathbf{x}^{H}
    =
    \mathbf{x}-\mathbf{x}^{L},
\end{equation}
where $K$ denotes the kernel size, and the superscripts $L$ and $H$ denote the low- and high-frequency components, respectively.

\begin{table*}[t]
\centering
\caption{Overall performance comparison on PTB-XL and Chapman-Shaoxing datasets.}
\label{tab:overall}
\setlength{\tabcolsep}{3pt}
\scriptsize
\begin{tabular*}{\textwidth}{
@{\extracolsep{\fill}}
l
cccc
@{\hspace{12pt}}
cccc
@{}
}

\toprule

& \multicolumn{4}{c}{\textbf{PTB-XL (in-domain)}}
& \multicolumn{4}{c}{\textbf{Chapman-Shaoxing (out-of-domain)}} \\

\cmidrule(r){1-1}
\cmidrule(lr){2-5}
\cmidrule(l){6-9}

Method
& MSE ($\times10^{-3}$)$\downarrow$
& CoS ($\times10^{-2}$)$\uparrow$
& SNR (dB)$\uparrow$
& MAD ($\times10^{-1}$)$\downarrow$
& MSE ($\times10^{-3}$)$\downarrow$
& CoS ($\times10^{-2}$)$\uparrow$
& SNR (dB)$\uparrow$
& MAD ($\times10^{-1}$)$\downarrow$ \\

\cmidrule(r){1-1}
\cmidrule(lr){2-5}
\cmidrule(l){6-9}

LI
& 7.477\rlap{$^{*}$}
& 90.211\rlap{$^{*}$}
& 8.592\rlap{$^{*}$}
& 8.773\rlap{$^{*}$}
& 4.289\rlap{$^{*}$}
& 95.114\rlap{$^{*}$}
& 10.549\rlap{$^{*}$}
& 11.114\rlap{$^{*}$} \\

SRECG~\cite{srecg}
& 0.422\rlap{$^{*}$}
& 99.367\rlap{$^{*}$}
& 19.751\rlap{$^{*}$}
& 3.711\rlap{$^{*}$}
& 1.227\rlap{$^{*}$}
& 98.729\rlap{$^{*}$}
& 17.535\rlap{$^{*}$}
& 6.592\rlap{$^{*}$} \\

DCAE-SR~\cite{dcaesr}
& 4.461\rlap{$^{*}$}
& 97.560\rlap{$^{*}$}
& 12.591\rlap{$^{*}$}
& 9.622\rlap{$^{*}$}
& 9.017\rlap{$^{*}$}
& 96.168\rlap{$^{*}$}
& 10.829\rlap{$^{*}$}
& 14.099\rlap{$^{*}$} \\

MSECG~\cite{msecg}
& 0.184\rlap{$^{*}$}
& 99.745\rlap{$^{*}$}
& 24.037\rlap{$^{*}$}
& 2.206\rlap{$^{*}$}
& 0.776\rlap{$^{*}$}
& 99.084\rlap{$^{*}$}
& 21.332\rlap{$^{*}$}
& 4.874\rlap{$^{*}$} \\

DiSR-ECG (10 shots)~\cite{disrecg} 
& 0.181\rlap{$^{*}$}
& 99.754\rlap{$^{*}$}
& 24.092\rlap{$^{*}$}
& \textbf{2.158}
& 0.642\rlap{$^{*}$}
& \textbf{99.376}
& 21.939\rlap{$^{*}$}
& 4.469\rlap{$^{*}$} \\

\cmidrule(r){1-1}
\cmidrule(lr){2-5}
\cmidrule(l){6-9}

DBFM (ours)
& \textbf{0.169}
& \textbf{99.759}
& \textbf{24.331}
&  2.162
& \textbf{0.631}
& 99.335
& \textbf{21.960}
& \textbf{4.375} \\

\bottomrule

\end{tabular*}

\vspace{1mm}

\begin{minipage}{\textwidth}
\footnotesize
$^{*}$ indicates a statistically significant difference from the proposed method
based on the paired Wilcoxon signed-rank test with Benjamini--Hochberg FDR
correction (adjusted  $p\text{-value}<0.05$).
\end{minipage}

\end{table*}

\subsection{Conditional flow matching}

\begin{figure}[t]
    \centering
    \includegraphics[width=\columnwidth]{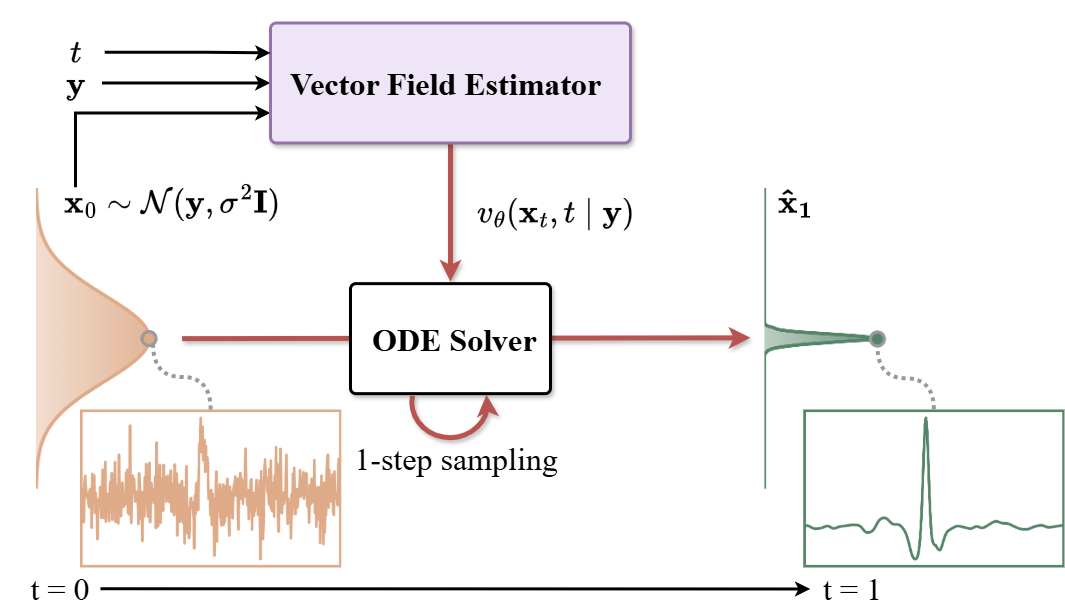}
    \caption{One-step inference of CFM for ECG super-resolution.}
    \label{fig:flowmatching}
\end{figure}

The formulation is applied independently to each frequency component $c \in \{L,H\}$. For brevity, we omit the component superscript $c$ in the following formulation, as it is identical across branches. Let $\mathbf{y}$ denote the LR observation, $\mathbf{x}_1$ the corresponding HR target, and $\mathbf{x}_0$ the source state.

Rather than adopting a data-independent Gaussian source, we construct the source distribution around the observed LR ECG. Generalized conditional flow matching allows flexible couplings between the source and target distributions~\cite{tong2023improving}. In our setting, the paired LR--HR observations provide a natural correspondence for constructing such a coupling. We define the conditional source distribution as
\begin{equation}
    p_0(\mathbf{x}_0\mid\mathbf{y})
    =
    \mathcal{N}\!\left(
        \mathbf{x}_0;
        \boldsymbol{\mu}_0(\mathbf{y}),
        \sigma_0(\mathbf{y})^2\mathbf{I}
    \right),
    \label{eq:source-dist}
\end{equation}
where $\boldsymbol{\mu}_0(\mathbf{y})=\mathbf{y}$ and
$\sigma_0(\mathbf{y})=\sigma$. The perturbation scale $\sigma$ is independently specified for each frequency component.
We adopt the conditional optimal transport (CondOT)~\cite{pmlr-v202-pooladian23a} probability path with $\mathbf{z}=(\mathbf{x}_0,\mathbf{x}_1)$. The corresponding linear transport path is
\begin{equation}
    \mathbf{x}_t
    =
    (1-t)\mathbf{x}_0+t\mathbf{x}_1,
    \qquad t\in[0,1],
    \label{eq:condot-path}
\end{equation}
with the conditional target vector field
\begin{equation}
    u_t(\mathbf{x}_t\mid\mathbf{z})
    =
    \frac{d\mathbf{x}_t}{dt}
    =
    \mathbf{x}_1-\mathbf{x}_0.
    \label{eq:condot-velocity}
\end{equation}
For a fixed source--target pair, the target velocity is therefore independent of $t$, while the state $\mathbf{x}_t$ varies along the path.

Following recent work on timestep sampling for single-step flow matching~\cite{yoshai2026caflow}, we employ a mixed time-sampling strategy that incorporates the initial state while retaining training along the transport path. Specifically, the timestep is sampled as
\begin{equation}
    t =
    \begin{cases}
        0, & \text{with probability } 0.5,\\
        u^r,\quad u\sim\mathcal{U}(0,1),
        & \text{with probability } 0.5,
    \end{cases}
    \qquad r=2.
    \label{eq:time-sampling}
\end{equation}
The resulting training objective uses an $L_1$ loss:
\begin{equation}
    \mathcal{L}_{\mathrm{CFM}}
    =
    \mathbb{E}_{t,q(\mathbf{z}),p_t(\mathbf{x}_t\mid\mathbf{z})}
    \left[
        \left\|
        v_\theta(\mathbf{x}_t,t\mid\mathbf{y})
        -
        u_t(\mathbf{x}_t\mid\mathbf{z})
        \right\|_1
    \right].
    \label{eq:proposed-cfm}
\end{equation}
The vector field estimator $v_\theta$ adopts the same network architecture as DiSR-ECG~\cite{disrecg}, conditioned on the corresponding LR component $\mathbf{y}$ and time $t$.

\subsection{One-step reconstruction}

As illustrated in Fig.~\ref{fig:flowmatching}, each frequency component is reconstructed from its LR-centered source state. Using a single Euler step from $t=0$ to $t=1$, the reconstructed HR component for $c\in\{L,H\}$ is given by
\begin{equation}
    \hat{\mathbf{x}}_1^c
    =
    \mathbf{x}_0^c
    +
    v_\theta^c(\mathbf{x}_0^c,0\mid\mathbf{y}^c),
    \label{eq:one-step}
\end{equation}
where $\mathbf{x}_0^c$ is sampled from the conditional source distribution in Eq.~\eqref{eq:source-dist}. The final HR ECG is obtained by combining the reconstructed low- and high-frequency components:
\begin{equation}
    \hat{\mathbf{x}}_1
    =
    \hat{\mathbf{x}}_1^L
    +
    \hat{\mathbf{x}}_1^H.
    \label{eq:reconstruction}
\end{equation}
DBFM therefore requires one vector-field evaluation per frequency component.

\section{Experiments}

\subsection{Datasets}

We use three publicly available ECG datasets for training and evaluation. The PTB-XL dataset~\cite{wagner2020ptb} serves as the primary dataset for model development and in-domain evaluation. It contains 21,799 ten-second 12-lead ECG recordings at 500~Hz from 18,869 patients.
The Chapman-Shaoxing dataset~\cite{PhysioNet-ecg-arrhythmia-1.0.0} is used exclusively for cross-dataset evaluation to assess the generalization ability of the proposed method to unseen data. It contains 45,152 ten-second 12-lead ECG recordings sampled at 500~Hz. No samples from this dataset are used during model training.
To improve robustness to real-world ECG artifacts, we incorporate noise recordings from the MIT-BIH Noise Stress Test Database (NSTDB)~\cite{moody1984bih} during training. NSTDB contains three common types of ECG artifacts: baseline wander (BW), muscle artifact (MA), and electrode motion (EM).

\subsection{Data preparation}

Each 10-s 12-lead ECG is sampled at 500~Hz and filtered using a second-order Butterworth bandpass filter between 1 and 45~Hz. The LR input is generated by downsampling the filtered signal by a factor of 10 to 50~Hz and linearly interpolating it back to 500~Hz, while the filtered signal serves as the HR target. For PTB-XL, we follow the official recommended split, using Folds 1--8 for training, Fold 9 for validation, and Fold 10 for testing~\cite{wagner2020ptb}.

During PTB-XL training, NSTDB noise is injected into the LR ECG with a probability of 0.5, using baseline wander, muscle artifact, or electrode motion noise at a type-specific SNR~\cite{hu2024lightweight}. Chapman-Shaoxing follows the same preprocessing without noise injection for cross-dataset evaluation.

\subsection{Evaluation metrics}

We evaluate ECG reconstruction quality using Mean Squared Error (MSE), Cosine Similarity (CoS), Signal-to-Noise Ratio (SNR), and Maximum Absolute Deviation (MAD). Lower MSE and MAD indicate smaller reconstruction errors, while higher CoS and SNR reflect better waveform similarity and signal fidelity, respectively. 

We further assess computational efficiency using the number of function evaluations (NFE), floating-point operations (GFLOPs), and real-time factor (RTF). NFE measures the number of model evaluations required for reconstruction, while GFLOPs quantify the computational cost of inference. RTF is calculated as the average inference time over 100 runs divided by the ECG duration, with lower values indicating faster processing.

{Statistical significance between DBFM and each baseline is assessed using a paired Wilcoxon signed-rank test at the record level, with Benjamini–Hochberg correction~\cite{benjamini1995controlling} applied across the four evaluation metrics for each baseline (adjusted \(p\text{-value}<0.05\)).}

\subsection{Implementation details}
The moving-average kernel size is set to $K=51$, with source noise scales $\sigma_L=0.05$ and $\sigma_H=0.2$. 
Both branches are trained independently for 400 epochs using the Adam optimizer~\cite{kingma2014adam}, with a batch size of 16 and an initial learning rate of $10^{-4}$. The learning rate is reduced by a factor of 0.1 every 150 epochs. At inference, each component is reconstructed using a single Euler step, resulting in two function evaluations in total. All experiments are conducted on an NVIDIA GeForce RTX 2080 Ti.

\subsection{Performance comparison with existing ECG SR methods}
We compare the proposed method with five ECG SR approaches, including linear interpolation (LI), SRECG~\cite{srecg}, DCAE-SR~\cite{dcaesr}, MSECG~\cite{msecg}, and DiSR-ECG~\cite{disrecg}. 

Table~\ref{tab:overall} summarizes the overall performance on the in-domain PTB-XL dataset and the out-of-domain Chapman-Shaoxing dataset.
On PTB-XL, DBFM achieves the best MSE, CoS, and SNR, while its MAD is slightly higher than that of DiSR-ECG, with no statistically significant difference. On Chapman-Shaoxing, DBFM achieves the best MSE, SNR, and MAD, while its CoS is slightly lower than that of DiSR-ECG, also with no statistically significant difference.

\begin{figure}[t]
    \centering
    \includegraphics[width=\columnwidth]{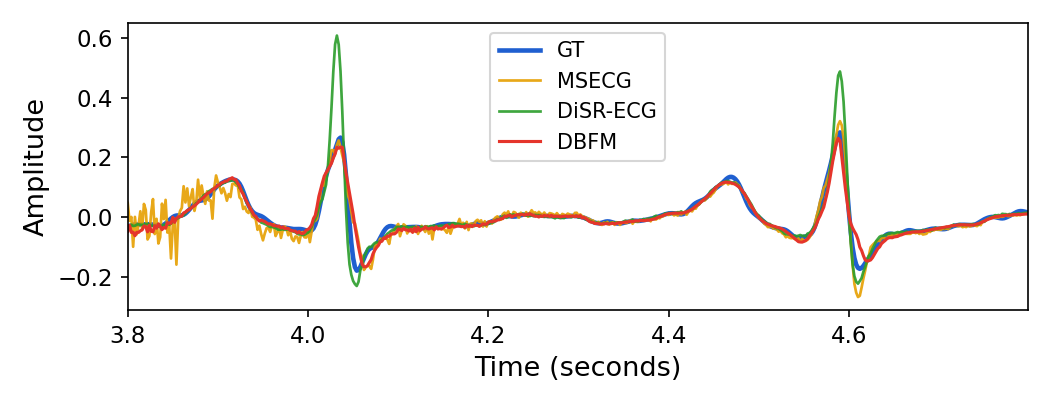}
    \caption{Qualitative comparison of ECG reconstruction on the Chapman-Shaoxing dataset.}
    \label{fig:qualitative}
\end{figure}

Fig.~\ref{fig:qualitative} shows the reconstructed ECG waveforms for record JS41952 (lead aVF, diagnosed with sinus tachycardia) from the Chapman-Shaoxing dataset. In this example, DBFM more closely follows the ground-truth (GT) waveform, particularly around the QRS complexes, whereas MSECG exhibits spurious high-frequency fluctuations and DiSR-ECG shows noticeable amplitude deviations near the QRS peaks. These observations further illustrate the ability of DBFM to preserve ECG waveform morphology with fewer reconstruction artifacts.

\begin{table}[t]
\centering
\caption{Computational efficiency comparison of ECG SR methods.}
\label{tab:efficiency}

\scriptsize
\setlength{\tabcolsep}{2pt}

\begin{tabular*}{\columnwidth}{
@{\extracolsep{\fill}}
lcccc
@{}
}
\toprule

Method
& Params (M)
& NFE$\downarrow$
& GFLOPs$\downarrow$
& RTF ($\times10^{-4}$)$\downarrow$ \\

\midrule

SRECG~\cite{srecg}
& 3.05
& 1
& 5.88
& $1.30{\pm}0.24$ \\

DCAE-SR~\cite{dcaesr}
& 28.33
& 1
& 69.75
& $6.99{\pm}0.25$ \\

MSECG~\cite{msecg}
& 1.91
& 1
& 2.01
& $15.00{\pm}0.41$ \\

DiSR-ECG (10 shots)~\cite{disrecg}
& 4.58
& 150
& 4958.60
& $4795.47{\pm}85.41$ \\

\midrule

DBFM (1 shot)
& 9.16
& 2
& 66.11
& $68.87{\pm}3.52$ \\

\bottomrule
\end{tabular*}

\end{table}
\begin{table}[t]
\centering
\caption{Ablation study on time-sampling strategies for training and the dual-band decomposition on PTB-XL.}
\label{tab:ablation}

\scriptsize
\setlength{\tabcolsep}{1.5pt}

\begin{tabular*}{\columnwidth}{
@{\extracolsep{\fill}}
l
cccc
@{}
}
\toprule

Variant
& MSE ($\times10^{-3}$)$\downarrow$
& CoS ($\times10^{-2}$)$\uparrow$
& SNR (dB)$\uparrow$
& MAD ($\times10^{-1}$)$\downarrow$ \\

\midrule

\multicolumn{5}{l}{\textit{Time Sampling}} \\

Uniform
& 0.240
& 99.649
& 22.445
& 2.792 \\

Uniform Mix
& 0.200
& 99.712
& 23.976
& 2.349 \\

Power
& 0.185
& 99.737
& 23.837
& 2.326 \\

Power Mix
& \textbf{0.169}
& \textbf{99.759}
& \textbf{24.331}
& \textbf{2.162} \\

\midrule

\multicolumn{5}{l}{\textit{Decomposition}} \\

w/o Decomp.
& 0.201
& 99.716
& 23.451
& 2.491 \\

w/ Decomp.
& \textbf{0.169}
& \textbf{99.759}
& \textbf{24.331}
& \textbf{2.162} \\

\bottomrule
\end{tabular*}

\end{table}

\subsection{Computational complexity analysis}

Table~\ref{tab:efficiency} compares the computational efficiency of neural-network-based ECG SR methods in terms of model parameters, NFE, GFLOPs, and RTF. Compared with DiSR-ECG, the strongest-performing baseline, DBFM substantially reduces both computational cost and inference time. Specifically, DBFM uses approximately 1/75 of the NFEs and GFLOPs of DiSR-ECG, while reducing the RTF by nearly 70-fold. This improvement comes at the cost of a twofold increase in model parameters, as DBFM employs separate models for the low- and high-frequency bands. This additional model complexity could potentially be reduced through asymmetric architectures for the two frequency bands or weight-sharing strategies. These directions may further narrow the efficiency gap between DBFM and lightweight direct-regression ECG SR methods, such as MSECG.

\subsection{Ablation study}

Table~\ref{tab:ablation} presents the ablation results for time sampling and dual-band decomposition. Both the mixed time-sampling strategy and dual-band decomposition contribute to the reconstruction performance of DBFM.

In terms of time sampling, Uniform uses $t\sim\mathcal{U}(0,1)$, Power uses $t=u^2$ with $u\sim\mathcal{U}(0,1)$, and their Mix variants assign $t=0$ with probability 0.5. Power sampling, which allocates more samples to the early region of the transport path, outperforms uniform sampling. Adding initial-state samples further improves both uniform and power sampling, with Power Mix achieving the best performance across all metrics.

For the dual-band decomposition ablation, removing the decomposition leads to consistent performance degradation across all metrics, supporting its effectiveness in reconstructing ECG components at different temporal scales.

\section{Conclusion}


In this paper, we proposed DBFM, an efficient dual-band flow-matching framework for ECG super-resolution. DBFM separately models low- and high-frequency ECG components and performs one-step reconstruction. Experiments on PTB-XL and Chapman-Shaoxing demonstrated strong performance and generalization to unseen data. Moreover, DBFM substantially reduces computational cost compared with iterative generative methods while preserving signal quality. These results highlight the effectiveness of DBFM for efficient ECG super-resolution. Future work will investigate DBFM on diverse real-world ECG signals and explore its applicability to downstream clinical tasks.

\vfill\pagebreak


\bibliographystyle{IEEEbib}
\bibliography{strings}

\end{document}